\documentclass[10pt,a4paper, onecolumn]{article}
\usepackage[utf8]{inputenc}
\usepackage{lineno}
\usepackage{a4wide}
\usepackage[top=60pt,bottom=60pt,left=60pt,right=60pt]{geometry}
\usepackage{multicol}
\usepackage{graphicx}
\usepackage{multirow}
\usepackage{amsmath,amssymb,amsfonts}
\usepackage{amsthm}
\usepackage{mathrsfs}
\usepackage[title]{appendix}
\usepackage{xcolor}
\usepackage{textcomp}
\usepackage{manyfoot}
\usepackage{booktabs}
\usepackage{algorithm}
\usepackage{algorithmicx}
\usepackage{algpseudocode}
\usepackage{listings}
\usepackage{authblk}
\usepackage{newfloat}
\usepackage{setspace}
\begin{document}

\title{Low-power scalable multilayer optoelectronic neural networks enabled with incoherent light}

\author[1,2*]{Alexander Song}
\author[1,2*]{Sai Nikhilesh Murty Kottapalli}
\author[1,2]{Rahul Goyal}
\author[3,4]{Bernhard Schölkopf}
\author[1,2]{Peer Fischer}

\affil[1]{Max Planck Institute for Medical Research, Heidelberg, Germany}
\affil[2]{Institute for Molecular Systems Engineering and Advanced Materials, Universität Heidelberg, Heidelberg, Germany}
\affil[3]{Max Planck Institute for Intelligent Systems, Tübingen, Germany}
\affil[4]{Department of Computer Science, ETH Zürich, Zürich, Switzerland}
\affil[*]{Equal contribution}

% \author{{Alexander Song}\inst{1,2}\thanks{\email{alexander.song@mr.mpg.de}} \textsuperscript{$\mp$} \and
%         \firstname{Sai Nikhilesh} \lastname{Murty Kottapalli}\inst{1,2}\thanks{\email{nikhilesh.kottapalli@mr.mpg.de}} \textsuperscript{$\mp$} \and
%         \firstname{Rahul} \lastname{Goyal}\inst{1,2} \and
%         \firstname{Bernhard} \lastname{Schölkopf}\inst{3,4} \and
%         \firstname{Peer} \lastname{Fischer}\inst{1,2}\thanks{\email{peer.fischer@mr.mpg.de}}
% }

\maketitle
          
% \date{October 2023}
\begin{abstract}
Optical approaches have made great strides towards the goal of high-speed, energy-efficient computing necessary for modern deep learning and AI applications. Read-in and read-out of data, however, limit the overall performance of existing approaches. This study introduces a multilayer optoelectronic computing framework that alternates between optical and optoelectronic layers to implement matrix-vector multiplications and rectified linear functions, respectively. Our framework is designed for real-time, parallelized operations, leveraging 2D arrays of LEDs and photodetectors connected via independent analog electronics. We experimentally demonstrate this approach using a system with a three-layer network with two hidden layers and operate it to recognize images from the MNIST database with a recognition accuracy of 92\% and classify classes from a nonlinear spiral data with 86\% accuracy. By implementing multiple layers of a deep neural network simultaneously, our approach significantly reduces the number of read-ins and read-outs required and paves the way for scalable optical accelerators requiring ultra low energy.
\end{abstract}
\doublespacing
\section{Introduction}

Deep learning is now ubiquitous for solving problems ranging from image recognition to drug discovery \cite{lecun2015deep}. Critical to this success is the use of ever larger deep learning models and datasets, that come with correspondingly rapid increases in required computing resources \cite{amodei2018ai,sevilla2022compute}. This increased demand \cite{desislavov2021compute,horowitz20141} has spurred research into alternative computing technology \cite{christensen20222022,caulfield2010future}. Research in optical computing has been explored for decades \cite{goodman1978fully,farhat1985optical,ohta1990optical,psaltis1988adaptive,denz2013optical}, and is currently undergoing a renaissance. The combination of the potentially dramatic energy savings \cite{caulfield2010future,miller2017attojoule} of light-based computation coupled with improvements in optoelectronics, photonics, and fabrication capabilities have led to promising first results \cite{wetzstein2020inference,shastri2021photonics}.\\

A major objective of contemporary optical computing approaches is to develop accelerators, energy-efficient implementations of small sections of modern neural networks \cite{hamerly2019large,bogaerts2020programmable}. Photonic accelerators make use of silicon fabrication to create a small number of high-speed, nonlinear photonic neurons \cite{tait2017neuromorphic,tait2019silicon,nahmias2019photonic} and recent implementations have reached computational power rivaling modern day GPUs \cite{feldmann2021parallel,xu202111,ashtiani2022chip}. Free-space accelerators typically have many more neurons at slower operating speeds and are potentially able to achieve even higher computation speeds \cite{lin2018all,hamerly2019large,miscuglio2020massively,spall2020fully,bernstein2021freely,zhou2021large,bernstein2022single,wang2022optical}.\\

Several challenges still need to be tackled before either photonic or free-space systems will be able to compete with existing computational hardware, such as system scalability, stability/accuracy, and interfacing with electronics \cite{christensen20222022}. One of the reasons for these challenges is the requirement of many systems for coherent light. Coherent systems enable complex summation \cite{lin2018all,zhang2021optical} and can make effective use of optical nonlinear activation functions \cite{zuo2019all,li2022all}. They typically require control over optical phase, resulting in strict requirements limiting system scale-up. Systems using amplitude-based computation in a free-space propagation setup \cite{chang2018hybrid,miscuglio2020massively,bernstein2021freely,wang2022optical} have primarily focused on using a single optical step between read-in and read-out of data and thus have not been extended to multilayer architectures (a recent example demonstrated a two-layer architecture \cite{wang2023image}). In these existing systems, the energy cost of electronic read-in/read-out constrains their overall efficiency. \\

\begin{figure}
    \centering
    \includegraphics[width=1.0\textwidth]{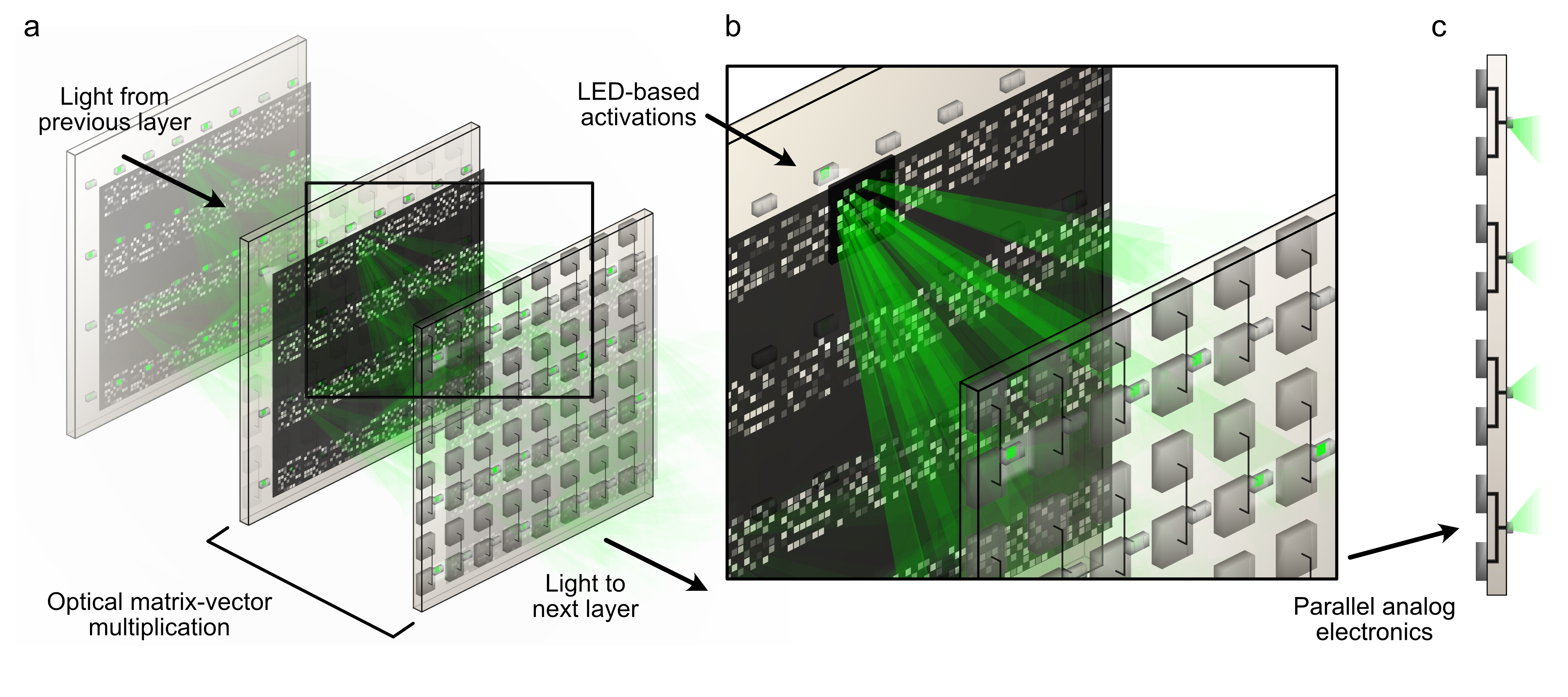}
    \caption{(a) The multilayer optoelectronic neural network uses a series of interleaved optical and electronic layers to implement matrix multiplication and nonlinearity, respectively. The inset illustrates (b) a nonnegative fully connected MVM that is implemented dynamically using a 2D array of incoherent light emitting diodes (LEDs), each encoding a neuron activation in our system. Each LED is associated with a 2D subarray of amplitude-encoded weights that map onto a 2D array of photodiodes (PDs). (c) An electronic board contains a parallel array of neurons each associated with a pair of photodiodes representing the positive and negative inputs to the neuron.}
    \label{fig:fig-1}
\end{figure}

In this work, we illustrate the potential of a multilayer incoherent optoelectronic accelerator. By deploying multiple optical interconnects with nonlinear activation functions between layer in a single system, the cost of electronic interfacing is greatly reduced, thereby opening the way for implementing scalable deep neural network architectures. We introduce and experimentally demonstrate a computing paradigm based on paired optoelectronic boards and optical interconnects, respectively describing nonlinear activation and weight matrix operations of a neural network (FIG \ref{fig:fig-1}). Our system builds upon and is smoothly extended by prior work implementing optoelectronic activation functions \cite{tait2017neuromorphic,williamson2019reprogrammable,pierangeli2021photonic,wang2023image} and matrix operations \cite{chang2018hybrid,shi2022loen,hamerly2019large,miscuglio2020massively,bernstein2021freely,miller2012all,matthes2019optical,kulce2021all,buddhiraju2021arbitrary}.\\

Our work is experimentally realized using off-the-shelf components on printed circuit boards and amplitude masks. The focus of the work is to demonstrate an optoelectronic computing paradigm that consists of individual units that can be straightforwardly scaled up in both the number of neurons and the number of layers. The system is designed so that networks previously trained on conventional computing hardware can be directly deployed onto the accelerator. This focus on inference-only systems is driven by the fact that roughly half of energy spent for AI currently goes into inference rather than training \cite{Patterson_2022}. Large models such as GPT-4 are trained for months on compute cluster containing tens of thousands of GPUs. While training takes place rarely, more than 100 million users place enormous demands on computing resources for inference \cite{Viswanathula_2023}. This greatly eases the ability for systems based on this paradigm to be adopted for industrial applications. These assets, in combination with advances in high-speed analog electronics, pave the way for large-scale implementations.\\

\section{Results}

\subsection{Multilayer optoelectronic neural network}\label{sec:results-system}
Modern neural network models commonly include a series of matrix-vector multiplications (MVM) and nonlinear activations. The matrix in these multiplications frequently take the form of either fully-connected matrices or convolution operations and the most commonly used nonlinear activation is the rectified linear (ReLU) function. 

\begin{figure}
    \centering
    \includegraphics{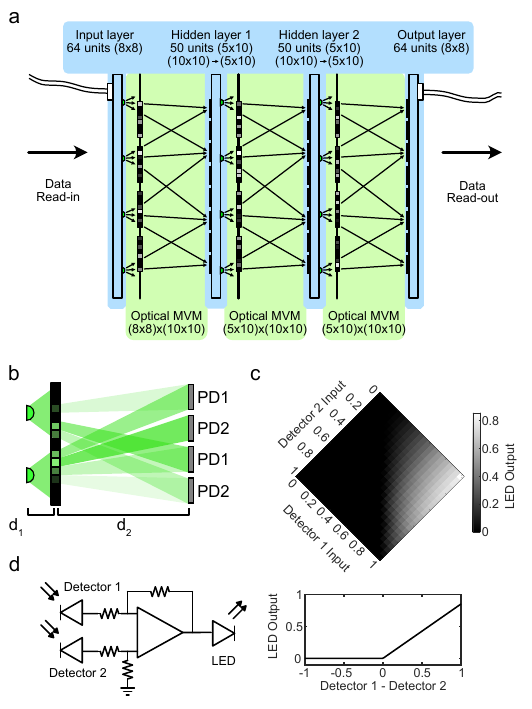}
    \caption{Schematic of our multi-layer optoelectronic neural network implementation with optical operations (green) and electronic operations (blue). (a) Data is read-in electronically to an Input layer with 64 units arranged on an 8x8 array of LEDs. A fully connected matrix-vector-multiplication (MVM) maps light from these units to a $10 \times 10$ array of photodidoes (PDs). Hidden layer 1 combines pairs of values from the PDs to drive a 5x10 array of LEDs. A second MVM and hidden layer implement Hidden layer 2 and a third MVM is mapped onto an 8x8 array of PDs of the Output Layer. (b) Ray-tracing illustrates how a fully-connected MVM operation is performed. (c) Amplitude weights are nonnegative, and a pair of photodiodes are fed into an analog electronic circuit that performs a differencing operation before driving an LED. (d) Example output LED response to a pair of detector inputs. Negative currents in the circuit are truncated by the LED, effectively implementing a linear rectification.}
    \label{fig:fig-2}
\end{figure}

Our experimental implementation of a multilayer optoelectronic neural network consists of four electronic boards representing an input layer, two hidden layers and an output layer (FIG. \ref{fig:fig-2}a) with optical MVMs in between. Free-space optics (green) execute a nonnegative fully-connected MVM while analog electronics (blue) perform differential photodetection, signal amplification, nonlinearity application, and light emission. 

While the setup implements fully connected MVMs, the 1D vectors of neuron activations in the input and hidden layers are mapped onto 2D arrays of light emitting diodes (LEDs). In this case of the input layer, a vector of 64 inputs is converted by an analog to digital converter (ADC) to the light intensity an $8 \times 8$ array of LEDs. Our approach uses the incoherent light from this LED array to perform the MVM in a lensless fashion using only nonnegative weights encoded on an amplitude mask (FIG. \ref{fig:fig-2}b, see Methods). Each LED positioned along the 2D array is associated with a 2D subarray of amplitude encoded weights that maps onto the 2D array of photodiodes (PDs) of the subsequent layer.

In our experiments, we use a liquid crystal display (LCD) to dynamically encode the amplitude mask. Other approaches, such as using phase-change materials \cite{feldmann2021parallel} or photomasks may also be used as passively encoded amplitude masks and improve the energy efficiency of system.

The positions and sizes of the weights on the amplitude mask are determined using ray-tracing (FIG. \ref{fig:fig-2}b). The mask is positioned at an axial distance $d_1$ away from the LED array and the PD array is positioned an additional distance $d_2$ further. This results in a magnification factor $M = \frac{d_1 + d_2}{d_1}$, which is both the size and shift scaling factor for the amplitude weights. This is used to determine the regions where light from $LED^i$ propagating towards $PD^j$ intersect the amplitude mask. This transmission of these regions are set to weight $W^{ij}$ for all $i,j$. We choose parameters for the LED die size, LED spacing, PD active area, PD spacing, and M to minimize crosstalk between the LED and PD pairs and the weights.

The system uses differential photodetection in the hidden layers to convert output values from the nonnegative MVM into a real valued MVM. A single neuron in a hidden layer has two PD inputs, corresponding to positive and negative portions of the neuron activation. These inputs are subtracted from each other using an operation amplifier (op-amp) differencing circuit. The circuit then amplifies the differenced input and drives an output LED. As an LED only emits light when forward biased, the circuit naturally implements a ReLU on the differenced input (FIG. \ref{fig:fig-2}c). For these experiments, we designed these circuits on printed circuit boards (PCBs) using commercially available integrated circuit chips (IC) and passive electronic components (see Methods, FIG. \ref{fig:fig-2}d).

The output from a hidden layer propagates through an optical MVM which may be used to drive another hidden layer. The process repeats until the output layer, which has a 2D PD array whose signal is read-out using an analog to digital converter (ADC) to a computer. The entire multilayer optoelectronic neural network runs continuously with sets of inputs and outputs synced to a clock.

\subsection{Image classification}\label{sec:results-mnist}
We tested the multilayer optoelectronic neural network by performing image classification on a downscaled version of the MNIST handwritten digit dataset. The dataset consists of $28 \times 28$ pixel images of handwritten digits between 0 and 9. We first downscale the digits to a $7 \times 7$ image and then pad the result with zeros and linearize it to form a length 64 vector. These linearized vector inputs were trained in PyTorch with a multilayer perceptron with the same network structure as our system (see Methods). Weights in the fully connected layers are constrained to experimentally determined maximum and minimum weights and experimentally determined offsets are added in the hidden layer differencing operation.

\begin{figure}
    \centering
    \includegraphics{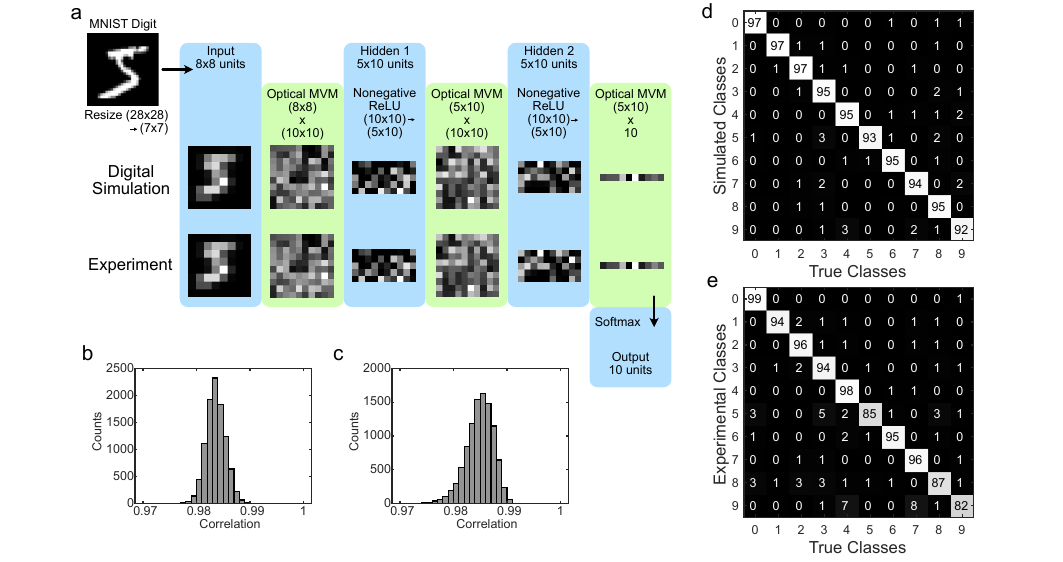}
    \caption{MNIST digit classification with a three-layer optoelectronic neural network. (a) Example propagation of a trained miniaturized MNIST digit through the three-layer network. Digital simulation values are compared to the analog experimental values. (b) Correlation between simulation and experiment of activations in Hidden layer 1 in response to individual miniaturized MNIST digits (c) Same as (b), but in Hidden layer 2. (d) Confusion matrix of estimated classes for simulated results, in percent. (e) Same as (d), but for experimental results}
    \label{fig:fig-3}
\end{figure}

After training, the weights are loaded onto the amplitude masks in the optical layers of our setup. For forward inference, the downscaled MNIST inputs are read in one at a time to our input board and propagated through the system. An example digit propagation through each of the layers as compared to the simulated values is shown in FIG. \ref{fig:fig-3}a. After each propagation, the outputs were digitized and fed back to the source board. Correlation between experimental values and digital simulation values of the neuronal activations in the hidden layers are high, demonstrating that errors due to cross-talk, nonlinearity in the LED response, and errors in the optical weight response are minimal (FIG. \ref{fig:fig-3}b,c).

For the task of classifying the MNIST handwritten digit dataset, this optoelectronic neural network attains a classification accuracy of $92.3\%$ in experiments as opposed to a classification accuracy of $95.4\%$ in the digital simulation (FIG. \ref{fig:fig-3}d,e). We followed up these experiments using the full multilayer opto-electronic neural network with all optical and electronic layers implemented simultaneously. In these classification experiments, we obtained an overall accuracy of $91.8\%$ with a test data simulation classification accuracy of $91.2\%$ and an experimental test data classification accuracy of $91.1\%$.

The protocol allows for a good alignment of each individual layer in the network with their corresponding optical weight masks giving a close match with simulations. This performance is in contrast to a digital classification accuracy of $82.4\%$ for a linear fully connected network in performing classification on the downscaled MNIST digits. This result demonstrates the advantages of the nonlinearity introduced in our network over the linear single layer performance.

\begin{figure}
    \centering
    \includegraphics{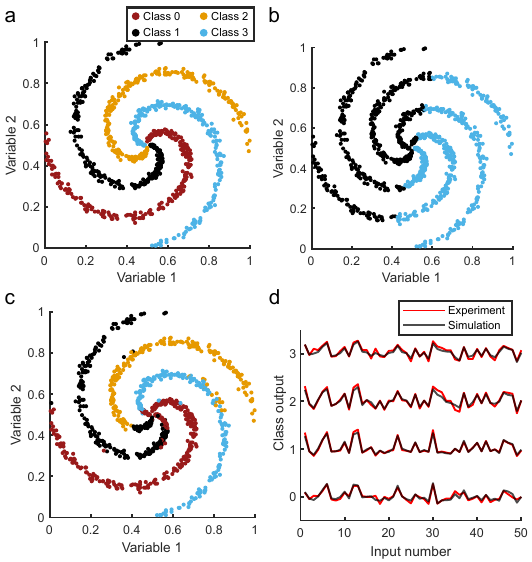}
    \caption{(a) A nonlinear four-class spiral data classification problem with two input variables. Each of the classes corresponds to one arm of the spiral. (b) The best linear classifier classifies this problem with an accuracy of 30.1\%. (c) Experimental classification using the multi-layer network as described in FIG. \ref{fig:fig-3}a obtains a classification accuracy of 86.0\%. (d) Comparison between simulation and experiment of the trained network output values.}
    \label{fig:fig-4}
\end{figure}

To further demonstrate the advantage of multiple nonlinear layers in the neural network, we setup a model of a two-input, four-class nonlinear spiral classification problem (FIG. \ref{fig:fig-4}). In this problem, a linear classifier has an accuracy of $30.1\%$, while the experimental output of our system is able to achieve a classification accuracy of $86.0\%$ (FIG. \ref{fig:fig-4}b,c). The direct outputs of the setup closely match the expected simulation results for the trained network (FIG. \ref{fig:fig-4}d) and the overall performance closely matches the best predicted performance.

\subsection{Deep optical accelerators with weight transfer}\label{sec:results-transfer}
Modern neural network architectures are large and complex, using dozens of layers with highly variable numbers of neurons and connections between layers. As such, it is impractical to completely replicate these architectures with optical/photonic approaches, including the multilayer optoelectronic neural network. A more useful application of these approaches is to implement reasonable portions of modern network architectures as an accelerator, especially if weights and structure from pre-trained networks can be directly transferred to the accelerator.

\begin{figure}
    \centering
    \includegraphics{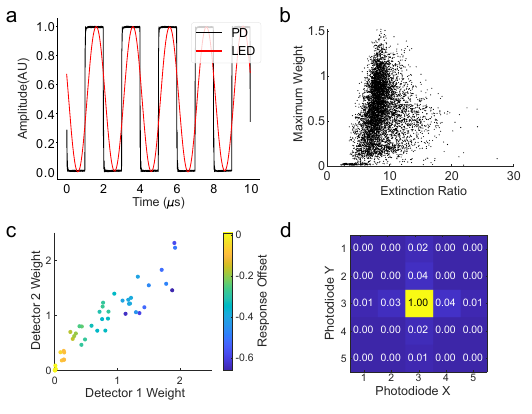}
    \caption{Calibration of the multilayer optoelectronic neural network (a) Temporal response of the electronics in response to changes in photodetector input. (b) Distribution of maximum optical weights and extinction ratio of an amplitude mask implemented with a liquid-crystal display (LCD). (c) Scatterplot of PD1 and PD2 from pairs of PDs implemented a nonnegative ReLU with a color-coded bias offset for individual pairs. (d) Average cross-talk distribution from weights implemented on the LCD.}
    \label{fig:fig-5}
\end{figure}

We have shown the multilayer optoelectronic neural network can flexibly implement some of the most common building blocks of modern neural networks, fully-connected MVMs and ReLUs. Additionally, these building blocks are high-speed (FIG. \ref{fig:fig-5}a) and independent, suggesting good future scaling for implementations with larger numbers of neurons. However, as our approach relies on analog computing, variability in the optical and electronic responses limits the performance of direct weight transfer. 

Measurements of our experimental implementation (FIG. \ref{fig:fig-5}b-d) show a moderate amount of variability in the weight response, LED brightness, and photodiode response, a result primarily due to the variability in the discrete commercial electronics used on the PCBs. In particular, a small number of neurons account for the majority of the variance. In this case, low-performance neurons may be excluded from individual layers, resulting in a remaining population of units that have a suitably uniform performance, a process analogous to the selection of individual cores on a microprocessor. The remaining variance in these properties can be normalized on the amplitude mask by elementwise multiplication of trained weights with inverse measured weight distribution during weight transfer.

Two additional properties that affect the performance of the multilayer optoelectronic neural network as an accelerator are crosstalk (FIG. \ref{fig:fig-5}d) and nonlinearity in the LED forward bias response. The measured crosstalk values are low, and do not substantially change the performance of our device during MNIST classification. Nonlinearity in the LED response similarly had a minimal effect on the performance and can be corrected with more sophisticated electronic circuits that are not reliant on op-amps for driving the LED response.

\begin{figure}[!hbt]
    \centering
    \includegraphics{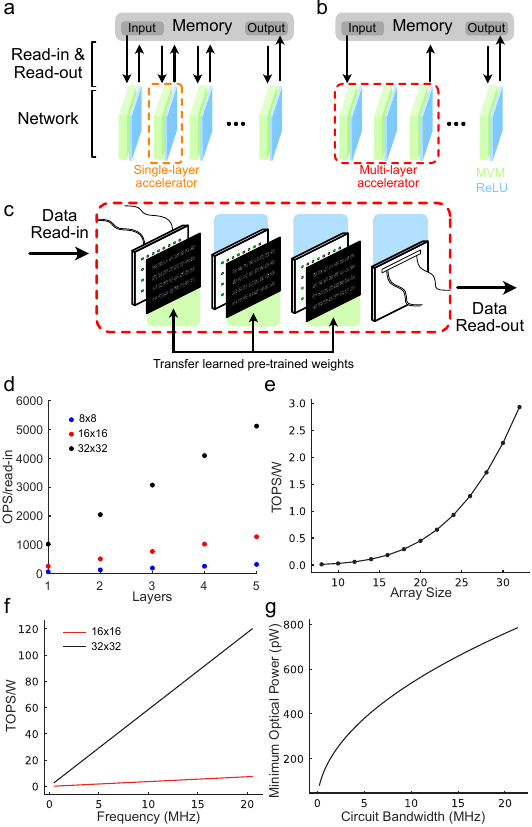}
    \caption{Advantage of the multi-layer optoelectronic neural network for neuromorphic computing. (a) Single-layer optoelectronic neural network accelerators (orange dotted box) require read-in and read-out of data to the accelerator for each layer. (b) A multi-layer accelerator (red dotted box) dynamically stores intermediate data, reducing the amount of data that must be read-in and read-out by a factor equal to the number of layers processed. (c) Our multi-layer accelerator implements three opto-electronic matrix-vector multiplications (MVM) each followed by a ReLU. This implementation can transfer over weights from pre-trained neural networks. (d) Number of compute operations performed per data read-in operation as a function of array size (colors) or layers (axis) in accelerator. (e) Scaling of system performance as a function of the photodiode array size in tera-operations per second (TOPS) per watt. (f) Scaling of  accelerator as the frequency of operation is varied (g) Minimum usable optical power in the system with varying circuit bandwidth.}
    \label{fig:fig-6}
\end{figure}

As an accelerator, one of the major advantages of our approach is its ability to implement multiple layers of a neural network simultaneously (FIG. \ref{fig:fig-6}a-c). One major bottleneck of conventional computing approaches is due to the von Neumann architecture where data is temporarily read-in and read-out of memory at each computation step. Optical/photonic accelerators that implement a single layer of a neural network suffer the same limitation and the energy cost of read-in/read-out of data dwarfs the energy cost of the computation itself \cite{wang2022optical}. Our approach, by implementing multiple layers simultaneously, reduces the read-in/read-out cost by a factor equal to the number of layers implemented (FIG. \ref{fig:fig-6}d), an advantage that grows with network depth. % \cite{Treiber_2023}.

As the system is designed to be energy efficient and scalable, we examine the power use per operation of the system as it is scaled up for practical implementations. In our proof-of-concept implementation with an $8 \times 8$ grid of photodiodes and a $4 \times8$ grid of LEDs operating at 500 kHz, we obtain a performance-per-watt of 11.61 GOPS/W. This is in contrast to a GPU from earlier generations such as NVIDIA 1080 which has a calculated performance-per-watt of about 49 GFLOPS/W or the A30 at 1 TFLOPS/W  \cite{desislavov2021compute}. Our system can be scaled both spatially and temporally. We examine the evolution of performance-per-watt of the system as both of those factors are scaled up. The energy performance of the system improves quadratically as the array sizes increases to $32 \times 32$ (FIG. \ref{fig:fig-6}e). beyond which diffraction effects become significant for moderate board sizes (see Methods). On scaling up to a $32 \times 32$ grid, the present approach yields a predicted performance of 2.92 TOPS/W. At higher speeds, the performance increases further, topping out at a prediced 120 TOPS/W (FIG. \ref{fig:fig-6}f,g), all the while using specifications from off-the-shelf components (see Methods).

\section{Discussion}
We have demonstrated a multilayer optoelectronic neural network based on interleaved optical and optoelectronic layers. The incoherent optical layers are simple, requiring only a single amplitude mask to perform fully connected MVMs. Similarly, the optoelectronic hidden layers rely on only basic electronic components, consisting of 2D arrays of photodetectors and LEDs connected locally by analog electronics. Our experimental setup with three MVMs and two hidden layers successfully classified handwritten digits, reaching a fidelity almost equal to values from digital simulation. Measurements of the response of individual neurons in each layer suggest this approach is suitable for direct transfer of weights from sections of modern neural networks architectures and used as a multi-layer optical accelerator for neural network inference.

We designed our system to be reminiscent of modern LED displays - where an LED array backlight projects through an LCD - combined with 2D photodiode arrays. These two components, when combined with local, independent analog processing, result in a computation platform that is suitable for large-scale implementations with very high data-processing rates. Modern LED displays are starting to make use of mini-LEDS and micro-LEDS \cite{huang2020mini}, paving the way for miniaturization. Recent improvements in CMOS chip design and analog electronics suggest that the required technology for large-scale implementation and manufacture are already available. 

Our approach is general and extensible in several directions. LEDs with different wavelengths can be used to encode either positive/negative weights or separate processing channels. Other modern neural network layers may be implemented. The optical MVM can be adapted for large scale convolution operations \cite{chang2018hybrid,shi2022loen,wang2023image} and beamsplitters may be used to implement skip layers. Analog electronics are straightforwardly adapted for pooling layers, other nonlinear responses, or encoded to add bias terms. We believe these advantages and extensibility will allow the multilayer optoelectronic neural network approach to rapidly translate into a useful optical accelerator for neural network inference while at the same time dramatically reducing the energy requirements of such computations.

\section{Methods}
\subsection{MNIST dataset and processing}\label{sec:methods-mnist}

The MNIST handwritten digit dataset \cite{deng2012mnist} was used to demonstrate the operation of our multilayer optoelectronic neural network. The MNIST handwritten digit dataset consists of 60,000 images of handwritten digits between $0$ and $9$. Each of the images is $28 \times 28$ pixels in size. For use in our system, we downscaled the image to $7 \times 7$ pixels using bilinear interpolation. The downscaled images were padded with zeros along each of the two dimensions to form an $8 \times 8$ pixel input to our system.
%\subsubsection{CIFAR dataset and processing (stacking and spreading inputs) 
%[Optional]}\label{sec:cifar-dataset-and-processing-stacking-and-spreading-inputs-optional}
\subsection{Control software}\label{sec:methods-control}
The control software for running the multilayer optoelectronic neural network is written in Python. The code for controlling the DAC (PXIe-6739) and ADC (NI PXIe-6355) instruments uses the NI-DAQMX Python package. The control pipeline consists of preloading preprocessed input data to the DAC and triggering simultaneous read-in and read-out of data. Data are synced via either the on-board clock or posthoc. The SLMs (Holoeye LC2012) used to control the amplitude masks are controlled in Python using OpenCV or the Holoeye SLM Display SDK. A CMOS camera (FLIR ORX-10G-71S7M) is controlled using the FLIR PySpin SDK.

\subsection{Network training}\label{sec:methods-training}\
Network training was performed using PyTorch on the downscaled MNIST dataset with a 5:1 split of data for training:testing. The downscaled MNIST digits are padded and linearized ($64 \times 1$) before being presented to the network. The network architecture is as depicted in FIG. \ref{fig:fig-1}a, equivalent to a fully-connected feed forward neural network with input size 64 followed by two hidden layers of size 50 (including ReLUs) and an output layer of size 64. Only 10 output units are used for the 10 MNIST classes, and a Softmax is applied to convert the outputs to probabilities.

Two custom layers are used to define the fully connected MVM and ReLU operations in a non-negative manner. The fully connected layer is implemented as a matrix vector multiplication of neuron activations of length $n$ with a nonnegative weight matrix $W$ of size $n \times 2m$ where $m$ is the number of units of the downstream layer. The weight matrix $W$ is clamped to experimentally determined minimum and maximum values from the process detailed in the alignment and calibration section of Methods. To increase robustness of the experimental network performance, an alternative version of this layer has been implemented during training to also include reshaping the output activations with a crosstalk matrix that has been randomly shifted by small subunit distances. The ReLU is implemented as a paired differencing operation where the $2m$ inputs are split into $m$ pairs of values that are subtracted from each other forming $m$ real-valued activations. An experimentally determined offset is added to these activations before a ReLU operation is applied. Similar to above, to increase robustness of the network, a random perturbation is sometimes applied to the neuronal activations and offsets during training. Training was performed using the Adam optimizer.
 
\subsection{Electronics design and operation}\label{sec:methods-electronics}

The optoelectronic neural network implements optical matrix-vector multiplications by mapping light from a 2D array of LEDs (Wurth 150040GS73220, Vishay VLMTG1400) to a 2D array of photodiodes or phototransistors (OSRAM SFH2704, SFH3710). Light-detection, signal processing and amplification, and light-emission are performed using analog electronics and optoelectronics on a printed circuit board (PCB). The circuits used are designed in LTSpice to meet AC and DC performance requirements at operation frequencies of up to 1MHz and are then tested on a matrix board. We then design PCBs from those circuits in KiCad 6 using components from standard libraries. %(SnapEDA and Ultra Librarian).

We design three types of PCBs each corresponding to the input, hidden and output layers. The input board reads in analog data using a National Instruments (NI) digital-to-analog converter (DAC) PXIe-6739. 64 analog voltage inputs are converted to current values to drive an $8 \times 8$ array of LEDs. A modified Howland current pump circuit design implemented with operational amplifiers (op-amp) is used to drive each LED independently. 

An intermediate board implements one of the hidden layers in our optoelectronic neural network. In our experimental setup it is composed of a $5 \times 10$ array of independent units that each perform three operations: photodetection, differencing and amplification, and light emission. In each unit, photodetection acquires signals from $2 \times 1$ photodiodes amplified with a transimpedance amplifier. These two signals are subtracted from each other using an op-amp based differential amplifier. A circuit converts this signal to a current to drive a LED. The activations of the hidden layer are encoded as the output intensity of the LEDs, which naturally rectifies any negative current output to zero output intensity.

The output board consists of a 2D array of photodiodes whose signals are each amplified and converted to a voltage with a transimpedance amplifier. These voltages are read-out using an analog to digital (ADC) converter NI PXIe-6355 to a computer. A CMOS camera is also used in experiments in place of a photodiode array for characterization of the optical response and calibration for optical alignment.

\subsection{Optics design and operation}\label{sec:methods-optics}
Our system executes a fully connected optical matrix-vector multiplication by mapping light from the 2D LED plane to the 2D photodetector plane with weights encoded a single amplitude mask. A grayscale amplitude mask implemented on a liquid crystal display is used to encode the optical weight matrix in the multilayer optoelectronic neural network. A transmissive spatial light modulator (SLM, Holoeye LC2012) is used in conjunction with a pair of polarizers for this purpose. The SLM has a resolution of $1024 \times 768$ with a pixel size of $36\mu m$ and is presented an image composed of a subarray of weights for each LED that are mapped to the photodetector plane of the following layer.

Light from the LED plane propagates a distance $d_1$ before impinging the amplitude mask and then propagates a further distance $d_2$ before interacting with the photodetector plane. The magnification $M = \frac{d_1 + d_2}{d_1}$ describes the scaling factor for the shift ($x_{Amp}^{ij}-x_{LED}^i$) between a LED position $ x_{LED}^i $ and a position on the amplitude mask $x_{Amp}^{ij}$ at the photodetector plane. The output position on the photodetector plane is then $x_{PD}^j= x_{LED}^i +M(x_{Amp}^{ij}- x_{LED}^i)$. We position each weight $W^{ij}$ at $x_{Amp}^{ij}$ for each pair $i,j$ to satisfy this relationship.
If we sum the intensity contribution from each individual $LED^i$ from the preceding PCB on a $PD^j$, we obtain a sum of the product of each of LED intensities $I_{\text{LED}}^{i}$ and optical weights $w^{(i,j)}$

\[
 O_{\text{PD}}^{j} = \sum_{i}I_{\text{LED}}^{i} \;\cdot\; w^{(i,j)} 
\]

Similarly, we can calculate the signal at each photodetector and represent it as a product of the values of the optical weights and LED signals. For an $8 \times 8$ array of photodetectors, we can represent the detected signal as

\[
O = \left[ \begin{array}{cccccccc} \sum_{i=1}^{64} I_{\text{LED}}^{i} \cdot w^{(i,1)} & \sum_{i=1}^{64} I_{\text{LED}}^{i} \cdot w^{(i,2)} & \dots & \sum_{i=1}^{64} I_{\text{LED}}^{i} \cdot w^{(i,8)} \\ \sum_{i=1}^{64} I_{\text{LED}}^{i} \cdot w^{(i,9)} & \sum_{i=1}^{64} I_{\text{LED}}^{i} \cdot w^{(i,10)} & \dots & \sum_{i=1}^{64} I_{\text{LED}}^{i} \cdot w^{(i,16)} \\ \vdots & \vdots & \ddots & \vdots \\ \sum_{i=1}^{64} I_{\text{LED}}^{i} \cdot w^{(i,57)} & \sum_{i=1}^{64} I_{\text{LED}}^{i} \cdot w^{(i,58)} & \dots & \sum_{i=1}^{64} I_{\text{LED}}^{i} \cdot w^{(i,64)} \end{array} \right]
\]

Which can be split into the input matrix and weight matrix as follows

\[
O = \left[ \begin{array}{ccc} I_{\text{LED}}^{1} & \dots & I_{\text{LED}}^{8} \\ \vdots & \ddots & \vdots \\ I_{\text{LED}}^{57} & \dots & I_{\text{LED}}^{64} \end{array} \right] \cdot \left[ \begin{array}{c} w^{(i, 1)} \\ \vdots \\ w^{(i, 64)} \end{array} \right] = \left[ \begin{array}{c} O^{1}_{\text{PD}} \\ \vdots \\ O^{1}_{\text{PD}} \end{array} \right]
\]

We use Monte Carlo raytracing to simulate the light distribution from the LED plane to the PD plane. These simulations are used to better predict the distribution of light on the PD plane caused by individual LED and weight positions due to the non-uniformity in LED light distribution and angle-dependent effects on the amplitude mask plane. Additionally, these simulations estimate the spread of light on the PD plane due to the finite size of the LED die and amplitude weights. 

A modified angular spectrum propagation that uses the averaged output of optical propagations with randomized input phases was used to estimate the effects of diffraction on the optical propagation for both the experimental parameters in the experiment and also for a larger scale, $32 \times 32$ sized array.
 
\subsection{Alignment and calibration of optics/electronics}\label{sec:methods-alignment}
PCBs are fastened to an optical table using $1/2$’’ posts (Thorlabs) and custom-designed 3D-printed parts. The 3D printed parts include holes for a 60mm cage. The 60mm cage and $1/4$’’ cage rods are used to precisely position and separate the PCBs with respect to one another. Soldering of LEDs and photodiodes is performed with reflow soldering with a component placement error of $\pm 100 \mu m$. 

Optical masks are displayed on a SLM with a pair of polarizers. Errors due to gross optical alignment and component placement are dynamically corrected. Alignment is performed layer by layer with intermediate outputs imaged onto a CMOS camera. We use custom code to iteratively shift positions of weights from idealized positions to optimize performance and minimize weight crosstalk. This is also used to calibrate the SLM transmission response.

\section{Acknowledgements}
The authors thank Dr. B. Miksch for his assitance in circuit design and layout. The authors also gratefully acknowledge Dr. B. Miksch, Dr. V. Volchkov and Mr. L. Schlieder for insightful thoughts and comments. This work was in part supported by the European Research Council under the ERC Advanced Grant Agreement HOLOMAN (No. 788296)."

\section{Author Information}

Alexander Song and Sai Nikhilesh Murty Kottapalli: these authors contributed equally to this work. 

% \subsection{Authors and Affiliations}
% \textbf{1 Max Planck Institute for Medical Research, Heidelberg, Germany}

% Alexander Song, Sai Nikhilesh Murty Kottapalli, Rahul Goyal, and Peer Fischer

% \textbf{2 Institute for Molecular Systems Engineering and Advanced Materials, Universität Heidelberg, Heidelberg, Germany}

% Alexander Song, Sai Nikhilesh Murty Kottapalli, Rahul Goyal, and Peer Fischer

% \textbf{3 Max Planck Institute for Intelligent Systems, Tübingen, Germany}

% Bernhard Schölkopf

% \textbf{4 Department of Computer Science, ETH Zürich, Zürich, Switzerland}

% Bernhard Schölkopf

\subsection{Author Contributions}
AS, SMNK, BS, and PF conceived the project. AS and SMNK developed the optical and optomechanic hardware. SNMK and RS designed the circuits and layout. AS trained the neural networks and wrote the experimental control code. AS and SMNK collected data and analyzed the results. AS, SNMK, and PF wrote the manuscript, and all authors provided comments and contributions.

\subsection{Corresponding Author}
Correspondence to Alexander Song (alexander.song at mr.mpg.de)
\bibliographystyle{naturemag}

\section{Disclosure Statement}
A.S., S.N.M.K., B.S. and P.F. are inventors on a patent application related to multilayer opto-electronic networks (EP23162917.1).

% \bibliography{bib.bib}

\end{document}